\documentclass[prb,twocolumn,showpacs,amsmath,amssymb,preprintnumbers,superscriptaddress]{revtex4}
\usepackage{dcolumn}
\usepackage{bm}
\usepackage{graphicx}
\usepackage{color}

\begin{document}

\title{Temperature evolution of magnetic and transport behavior in 5\textit{d} Mott insulator Sr$_2$IrO$_4$: Significance of magneto-structural coupling}

\author{Imtiaz Noor Bhatti}\affiliation{School of Physical Sciences, Jawaharlal Nehru University, New Delhi - 110067, India.}
\author{R. Rawat}\affiliation{UGC-DAE Consortium for Scientific Research, University Campus, Khandwa Road, Indore - 452001, M.P, India.}
\author{A. Banerjee}\affiliation{UGC-DAE Consortium for Scientific Research, University Campus, Khandwa Road, Indore - 452001, M.P, India.}
\author{A. K. Pramanik}\email{akpramanik@mail.jnu.ac.in}\affiliation{School of Physical Sciences, Jawaharlal Nehru University, New Delhi - 110067, India.}

\begin{abstract}
We have investigated the temperature evolution of magnetism and its interrelation with structural parameters in perovskite-based layered compound Sr$_2$IrO$_4$, which is believed to be a $J_{eff}$ = 1/2 Mott insulator. The structural distortion plays an important role in this material which induces a weak ferromagnetism in otherwise antiferromagnetically ordered magnetic state with transition temperature around 240 K. Interestingly, at low temperature below around 100 K, a change in magnetic moment has been observed. Temperature dependent x-ray diffraction measurements show sudden changes in structural parameters  around 100 K are responsible for this. Resistivity measurements show insulating behavior throughout the temperature range across the magnetic phase transition. The electronic transport can be described with Mott's two-dimensional variable range hopping (VRH) mechanism, however, three different temperature ranges are found for VRH, which is a result of varying localization length with temperature. A negative magnetoresistance (MR) has been observed at all temperatures in contrast to positive behavior generally observed in strongly spin-orbit coupled materials. The quadratic field dependence of MR implies a relevance of a quantum interference effect.  
\end{abstract}


\maketitle
\section {Introduction}
The Ir-based 5\textit{d} transition metal oxides (TMOs) have been the topic of intense research in recent times due to their unusual properties.\cite{kim1,kim2,balent,naka,shapiro,yogesh,okamoto,moon,jack,senthil,yang,bahr} With increasing \textit{d} character in 5\textit{d} TMOs, the orbitals are spatially extended which substantially reduces the electronic correlation effect (\textit{U}). On the other hand, extended orbitals contribute in enhanced bonding interaction and orbital dependent physical phenomenon such as, orbital ordering, unusual magnetic properties, etc. Apart from these, the spin orbit coupling (SOC) effect becomes very prominent in these materials due to high value of atomic number of 5\textit{d} transition metals. This SOC effect plays an additional key factor in governing their detailed physical properties. With this reduced $\textit{U}$, the 5$\textit{d}$ TMOs are expected to be more metallic than its 3\textit{d} and 4\textit{d} counterparts. Surprisingly, however, they happen to be strong insulator. It is commonly believed that the strong SOC in iridium oxides causes splitting of $t_{2g}$ band into $J_{eff}$ = 1/2 and $J_{eff}$ = 3/2 multiples.\cite{kim1,kim2} Since Ir$^{4+}$ with five \textit{d} electrons adopts a low spin state, four of which fill the lower $J_{eff}$ = 3/2 while one partially fills the upper $J_{eff}$ = 1/2. The presence of \textit{U}, though small, further splits the narrow $J_{eff}$ = 1/2 band opening a Mott type gap which renders an insulating behavior. Nonetheless, the delicate competition among the various interactions such as, crystal field effect (CEF), SOC and \textit{U} plays a crucial role in realizing the $J_{eff}$ = 1/2 intriguing physics in these materials.

The perovskite-based layered compound Sr$_2$IrO$_4$ with celebrated K$_2$NiF$_4$ structure is an interesting member of 5\textit{d} TMOs. The Sr$_2$IrO$_4$ crystallizes in tetragonal structure with \textit{I4$_1$/acd} space group where the IrO$_6$ octahedra exhibits a rotation of about 11$^o$ around $c$ axis (see Fig. 1).\cite{crawford} This structural organization places an important interest on this material. For instance, the isostructural nature of Sr$_2$IrO$_4$ with 3\textit{d}- and 4\textit{d}-based superconducting materials La$_2$CuO$_4$ and Sr$_2$RuO$_4$, respectively has induced a special interest in search of superconductivity in this material.\cite{senthil,yang} Moreover, the octahedral distortion (rotation) is believed to bring in a Dzyaloshinsky-Moriya (DM) type antisymmetric exchange interaction which leads to a transition to weak ferromagnetic (FM) phase with transition temperature ($T_c$) $\sim$ 240 K. This material otherwise would show a collinear type antiferromagnetic (AFM) ordering.\cite{crawford,ye} However, at low temperature ($T$) below $\sim$ 100 K, a change in magnetization ($M$) is observed,\cite{ge,korneta,chikara} which appears not due to magnetic phase transition. In the present scenario of canted AFM spin ordering the change in moment could be related to spin ordering of Ir ions. In fact, a recent theoretical investigation has put forwarded a interrelation between spin canting angle and structural parameters in Sr$_2$IrO$_4$.\cite{jack} These observations emphasize for a temperature dependent structural analysis to comprehend the temperature evolution magnetic behavior. It can be mentioned that temperature dependent structural studies employing x-ray and neutron diffraction have been performed for Sr$_2$IrO$_4$, however, the detail analysis of data as well as finding a correlation between structural and magnetic behavior has not been attempted.\cite{crawford,ge1} Moreover, given an exotic insulating phase in present material the electronic transport mechanism needs to be investigated throughly including in presence of magnetic field ($H$).

We have investigated the interrelation between temperature dependent magnetic behavior and structural parameters by means of temperature dependent x-ray diffraction (XRD) measurements. The measured magnetization data show a transition to weak ferromagnetic phase around 238 K, however, at low temperature below around 95 K we find $M(T)$ show a sharp fall with lowering the temperature. Analysis of temperature dependent XRD data shows a sudden change in structural parameters around this temperature is responsible for this behavior. Resistivity ($\rho$) data show an insulating behavior where it rises drastically below 50 K. The mode of electronic transport has been understood to follow Mott's variable range hopping mechanism at three different temperature ranges. The quadratic nature of field dependence of magnetoresistance (MR) at low temperature implies the significance of quantum interference effect in this material.

\section {Experimental Details}
\begin{figure}
	\centering
		\includegraphics[width=8cm]{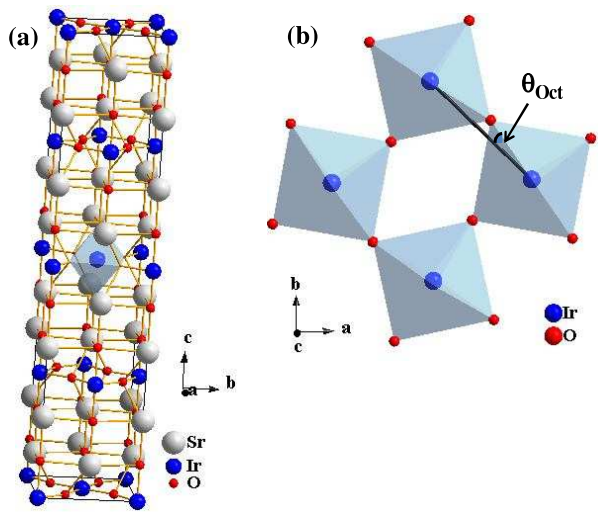}
	\caption{(Color online) (a) Schematic of Sr$_2$IrO$_4$ unit cell showing structural organization of IrO$_2$ and SrO layers. (b) The IrO$_2$ layer of Sr$_2$IrO$_4$ structure showing respective rotation of IrO$_6$ octahedra where $\theta_{Oct}$ is the angel of octahedral rotation around $c$ axis.}
	\label{fig:Fig1}
\end{figure}

Polycrystalline sample of Sr$_2$IrO$_4$ has been prepared using standard solid state ceramic method. The ingredient powder materials SrCO$_3$ and IrO$_2$ with phase purity more than 99.99\% (Sigma-Aldrich) are taken in stoichiometric ratio and ground well. The fine powders are calcined in air at 900$^o$C for 24 hours with heating and cooling rate of 3$^o$C/min. Calcinated powders are then palletized and sintered at 1000$^o$C and 1100$^o$C for 24 hours at same heating and cooling rates with an intermediate grinding. The phase purity of the sample has been checked using powder x-ray diffraction (XRD) with a Rigaku MiniFlex diffractomer with CuK$_\alpha$ radiation at room temperature. The reitveld analysis of XRD data shows the sample is in single phase withous any chemical impurity, and crystallizes in tetragonal phase with \textit{I4$_1$/acd} symmetry. The temperature dependent XRD measurements have been performed using a PANalytical X'Pert powder diffractomer in the temperature range of 298 - 20 K. A helium close cycle refregeretor (CCR) based cold head is used to achieve low temperature. Care has been taken for proper temperature stabilization by giving sufficient wait time before collecting data. Data have been collected in the 2$\theta$ range of 10 - 90$^o$ at a step of $\Delta 2\theta$ = 0.033$^o$ and a scan rate of 2$^o$/min. XRD data have been analyzed using Reitveld refinement program (FULLPROF) by Young \textit{et al}.\cite{young} DC magnetization data have been collected using a vibrating sample magnetometer (PPMS, Quantum Design) and electrical transport properties have been measured using a home-built insert fitted with Oxford superconducting magnet.

\section{Results and Discussions}
\subsection{Magnetization study}
Fig. 2 shows temperature dependent magnetization data for Sr$_2$IrO$_4$ measured in applied magnetic field of 10 kOe following zero field cooling (ZFC) and field cooled (FC) protocol. With lowering in temperature, $M(T)$ shows sharp increase around 240 K which is marked by paramagnetic (PM) to FM transition in this material as previously been observed.\cite{crawford} We find FM transition temperature $T_c$ $\sim$ 238 K which is estimated from inflection point in d$M$/d$T$ plot. The obtained $T_c$ is in agreement with reported studies of this material.\cite{crawford,cao,kini} On further lowering in temperature there is a bifurcation between ZFC and FC branches of magnetization data. A close observation reveals that bifurcation starts around 150 K. It is, however, interesting to notice that while $M_{FC}$ monotonically increases with a small slope, the $M_{ZFC}$ shows a steep decrease below 95 K ($T_M$) which continues down to lowest measuring temperature. This fall in moment below around 100 K has also been observed in case of both single and polycrystalline samples,\cite{ge,korneta,chikara} and may be the intrinsic to this material. There may be many underlying mechanism behind this behavior, yet the most obvious one could be structural phase/parameters change around this temperature. In fact, a theoretical calculation has predicted a interrelation between the rotation of IrO$_6$ octahedra and the spin canting angle.\cite{jack} This implies that a detailed structural analysis with temperature is required to understand the low-$T$ magnetic behavior.

\begin{figure}
	\centering
		\includegraphics[width=8cm]{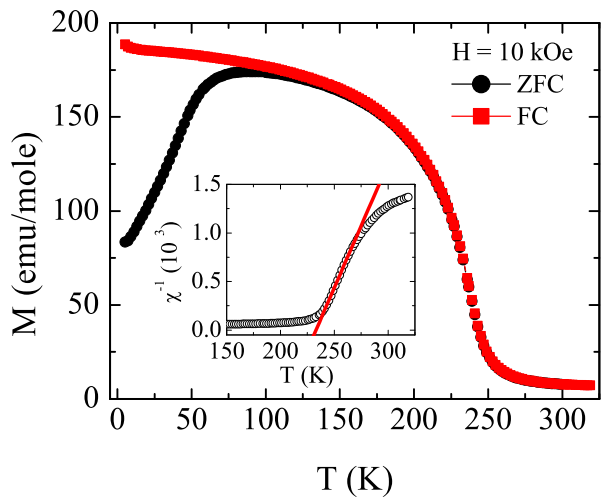}
	\caption{(Color online) DC magnetization data measured in 10 kOe applied field has been plotted as a function of temperature for Sr$_2$IrO$_4$. Data are collected following ZFC and FC protocol. Inset shows temperature dependent inverse susceptibility ($\chi^{-1}$ = $(M/H)^{-1}$) deduced from ZFC data for Sr$_2$IrO$_4$.}
	\label{fig:Fig2}
\end{figure}

The inset of Fig. 2 shows inverse susceptibility ($\chi^{-1}$), deduced from dc magnetization data $(M_{ZFC}/H)^{-1}$, as a function of temperature. Above $T_c$, the $\chi^{-1}(T)$ shows a linear behavior over a limited temperature range up to around 270 K following a Curie-Weiss behavior $\chi$ = $M/H$ = $C/(T - \theta_P)$, where $C$ and $\theta_P$ are the Curie constant and Curie temperature, respectively. From the straight line fitting (inset of Fig. 2) we estimate $\theta_P$ = 233.02 K. The value of $\theta_P$ not only very close to estimated $T_c$ of this material also it suggests spin ordering is of FM nature. Moreover, using the fitted parameters, we have calculated the effective PM moment ($\mu_{eff}$) in terms of Bohr magneton per formula unit as 0.56 $\mu_B$/f.u. The estimated $\mu_{eff}$ agrees with the other studies,\cite{cao,kini} and also the value matches with the expected one $\left(\mu_{eff} = g_J\sqrt{J_{eff}(J_{eff}+1)} \mu_B\right)$ for this material with single unpaired 5$d$ electron residing in $J_{eff}$ = 1/2 state. With $g_J$ = 2/3, the expected value of $\mu_{eff}$ comes as 0.57 $\mu_B$/f.u.   
 
To understand the magnetic field induced magnetic behavior in Sr$_2$IrO$_4$ with canted AFM-type spin ordering, we have recorded magnetic hysteresis loop $M(H)$ at 5 K (Fig. 3a) with field range of $\pm$70 kOe. At the highest applied field of 70 kOe, magnetization does not saturate, instead increases monotonically yielding a moment $\mu_H$ = 0.05 $\mu_B$/f.u. The measured $\mu_H$ is comparable with other studies of this material.\cite{crawford,ge,calder,cao,kini} This measured $\mu_H$ is significantly low compared to the expected value ($\mu_H$ = $g_JJ_{eff}\mu_B$) for this SOC dominated 5$d$ oxide material Sr$_2$IrO$_4$ which can be calculated as 1/3 $\mu_B$/f.u. This disagreement between the measured and expected values of $\mu_H$ is not very clearly understood. Given that there is strong coupling between structural and magnetic ordering in this material the structural investigation in presence of magnetic field and/or investigation introducing chemical doing would shed a light to understand this unusual behavior. On the other hand, this reduced moment has been explained from itinerant based band FM model using Rhodes-Wohlfarth plot:\cite{rhodes} with the ratio, $\mu_{eff}/\mu_H$ $>$ 1 the low moment in Sr$_2$IrO$_4$ has been ascribed due to itinerant FM behavior.\cite{cao} However, very high field measurement of moment is required to check the saturation moment before assigning it to band ferromagnetism. Note, that a reasonably high value of remnant magnetization $M_{rem}$ = 0.024 $\mu_B$/f.u. and coercive field $H_c$ = 9360 Oe (Fig. 3a) are obtained which is suggestive of high anisotropy present in this material.

\begin{figure}
	\centering
		\includegraphics[width=8.5cm]{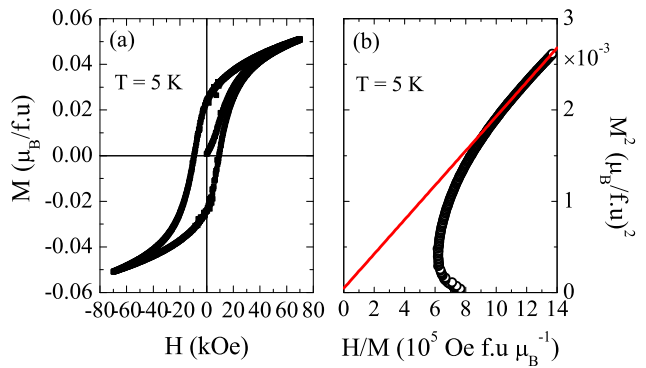}
	\caption{(Color online) (a) Magnetic field dependent magnetization is shown for Sr$_2$IrO$_4$ at 5 K. (b) Arrott plot ($M^2$ vs $H/M$) of magnetization data is shown at 5 K.}
	\label{fig:Fig3}
\end{figure}

To examine the nature of ground state magnetism we have plotted the virgin plot of $M(H)$ in form of Arrott plot:\cite{arrott} $M^2$ vs $H/M$. The Arrott plot has been successfully used to understand the magnetic state as the intercept on the positive $M^2$ axis due to an extrapolation of high field data in Arrott plot suggest a presence of spontaneous magnetization ($M_s$) in the system. Fig. 3b shows such Arrott plot related to 5 K $M(H)$ plot in Fig. 3a, where it can be seen a positive intercept on $M^2$ axis implying a FM ordering. From intercept the $M_s$ has been calculated to be 6.85 $\times$ 10$^{-3}$ $\mu_B$/f.u. which indicates the ground state of Sr$_2$IrO$_4$ is weak FM in nature.
         
\subsection{Temperature dependent XRD studies}
As there has been strong interplay between the structure and magnetic behavior in Sr$_2$IrO$_4$, therefore to understand the evolution of magnetism with temperature we have undertaken a detailed temperature dependent structural studies by means of XRD measurements. The data have been collected at 298, 250, 230, 200, 150, 100, 80, 60, 40, and 20 K encompassing the PM, FM and low temperature magnetic state below $T_M$. Fig. 4 shows XRD pattern at 298 K along with Rietveld refinement fitting which suggests good quality of sample without any chemically impure phase. The refinement result suggests that the sample crystallizes in tetragonal structure with \textit{I4$_1$/acd} symmetry.\cite{crawford} Details of structural parameters and crystallographic positions obtained from Reitveld analysis of XRD data at 298 K have been summarized in Table I. The ratio $R_{wp}/R_{exp}$, which is known to \textit{goodness of fit (GOF)}, is found to be around 1.38 showing a reasonably good fitting. The Ir-O-Ir bond lengths and bond angles, calculated using unit cell parameters and crystallographic positions have been shown in Table II. The IrO$_6$ octahedra has two distinct oxygen positions [two apical (O1) and four basal (O2)] where it is found to be slightly apically elongated (Table II). The $<$Ir-O1-Ir$>$ is found to be 180$^o$ showing they are not distorted along $c$ axis. The refinement, however, shows IrO$_6$ octahedra are rotated around $c$ axis by angle 11.26$^o$ which in conformity with previous reports (see Fig. 1).\cite{crawford}

\begin{figure}
	\centering
		\includegraphics[width=8cm]{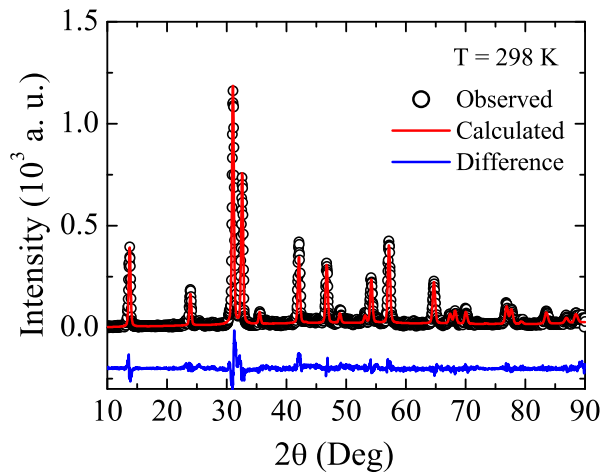}
		\caption{(Color online) XRD pattern collected at 298 K for Sr$_2$IrO$_4$ has been shown along with the Reitveld refinement. The refinement shows the material crystallizes in tetragonal structure with \textit{I4$_1$/acd} symmetry.}
	\label{fig:Fig4}
\end{figure}

\begin{table}[b]
\caption{\label{label} Structural parameters and crystallographic positions determined from the Rietveld profile refinement of the powder XRD patterns for Sr$_2$IrO$_4$ at 298 K. Here O1 refers to the apical oxygen, O2 refers to the basal oxygen which lies in the plane of the perovskite layer.}
\begin{ruledtabular}
\begin{tabular}{ccc}
Parameters &298 K\\
          
          &\textit{I4$_1$/acd}\\
\hline
a (\AA) &5.4980(2)\\
c (\AA) &25.779(1)\\
V (\AA${^3}$) &779.27(6)\\
Sr site &16d\\
x &0.0\\
y &0.0\\
z &0.17506(2)\\
Ir site &8a\\
x &0.0\\
y &0.0\\
z &0.0\\
O1 site &16d\\
x &0.0\\
y &0.0\\
z &0.07995(3)\\
O2 site &16f\\
x &0.20021(5)\\
y &0.20021(5)\\
z &0.25000\\
$R_{wp}$ &23.6\\
$R_{exp}$ &17.0\\ 
\end{tabular}
\end{ruledtabular}
\end{table} 

In Fig. 5, we have shown representative XRD pattern at three selected temperatures i.e., 298, 200 and 20 K which lies well within PM, FM and below $T_M$ magnetic state, respectively. It is evident in figure that there is no any major modification in patterns such as, splitting of peaks, asymmetric peak height, etc. A close observation shows there is a minute shift in peak positions. This is in favor of that while going to low temperature there is no structural phase transition, though there may be some changes in lattice parameters. Indeed, our Reitveld analysis of XRD data supports this observation (discussed below).

\begin{figure}
	\centering
		\includegraphics[width=8cm]{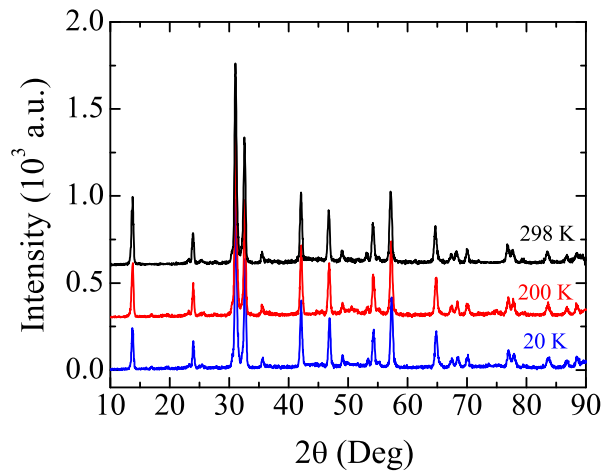}
	\caption{(Color online) XRD pattern have been shown for Sr$_2$IrO$_4$ at temperatures 298, 200 and 20 K. The pattern at 200 and 298 K have been shifted vertically for clarity.}
	\label{fig:Fig5}
\end{figure}
 
\begin{table}[h]
\caption{\label{label} Ir-O-Ir bond length ($d$) in \AA, bond angle ($<>$) in Deg are given for Sr$_2$IrO$_4$ at 298 K. Here O1 refers to the apical oxygen, O2 refers to the basal oxygen which lies in the plane of the perovskite layer.}
\begin{ruledtabular}
\begin{tabular}{ccc}
Parameters &298 K\\
          
          &\textit{I4$_1$/acd}\\
\hline
d$_{Ir-O1}$ (\AA) &2.0610(4)\\
d$_{Ir-O2}$ (\AA) &1.9820(3)\\
$<$Ir-O1-Ir$>$ (Deg) &180\\
$<$Ir-O2-Ir$>$ (Deg) &157.47\\
$\theta_{Oct}$ (Deg) &11.26\\
\end{tabular}
\end{ruledtabular}
\end{table}

The temperature dependent lattice parameters ($a$ and $c$) refined with Rietveld analysis are shown in Figs. 6a and 6b. Figs. 6c and 6d, respectively show the $c/a$ ratio and unit cell volume ($V$) as a function of temperature. Fig. 6 also shows two vertical dotted lines representing $T_c$ and $T_M$ of Sr$_2$IrO$_4$. It can be observed in figure that $a$ decreases, $c$ increases, $c/a$ increases and $V$ decreases with lowering temperature. The increase of parameter $c/a$ implies tetragonal structure of this compound is further elongated along $c$ axis. However, temperature evolution of parameters is not monotonous. With lowering temperature, all the parameters exhibit a change in slope at onset of FM phase transition i.e., around $T_c$. Within FM phase, although the parameters change with temperature but their rate of change is almost maintained. Interestingly, around $T_M$ all the parameters show a sudden drastic change, and with further decrease in temperature parameters change with slower rate toward saturation. We do not, however, find any structural phase change cooling across $T_c$ and $T_M$ (Fig. 5). These changes of structural parameters around $T_c$ and $T_M$ is quite intriguing.  

\begin{figure}
	\centering
		\includegraphics[width=7cm]{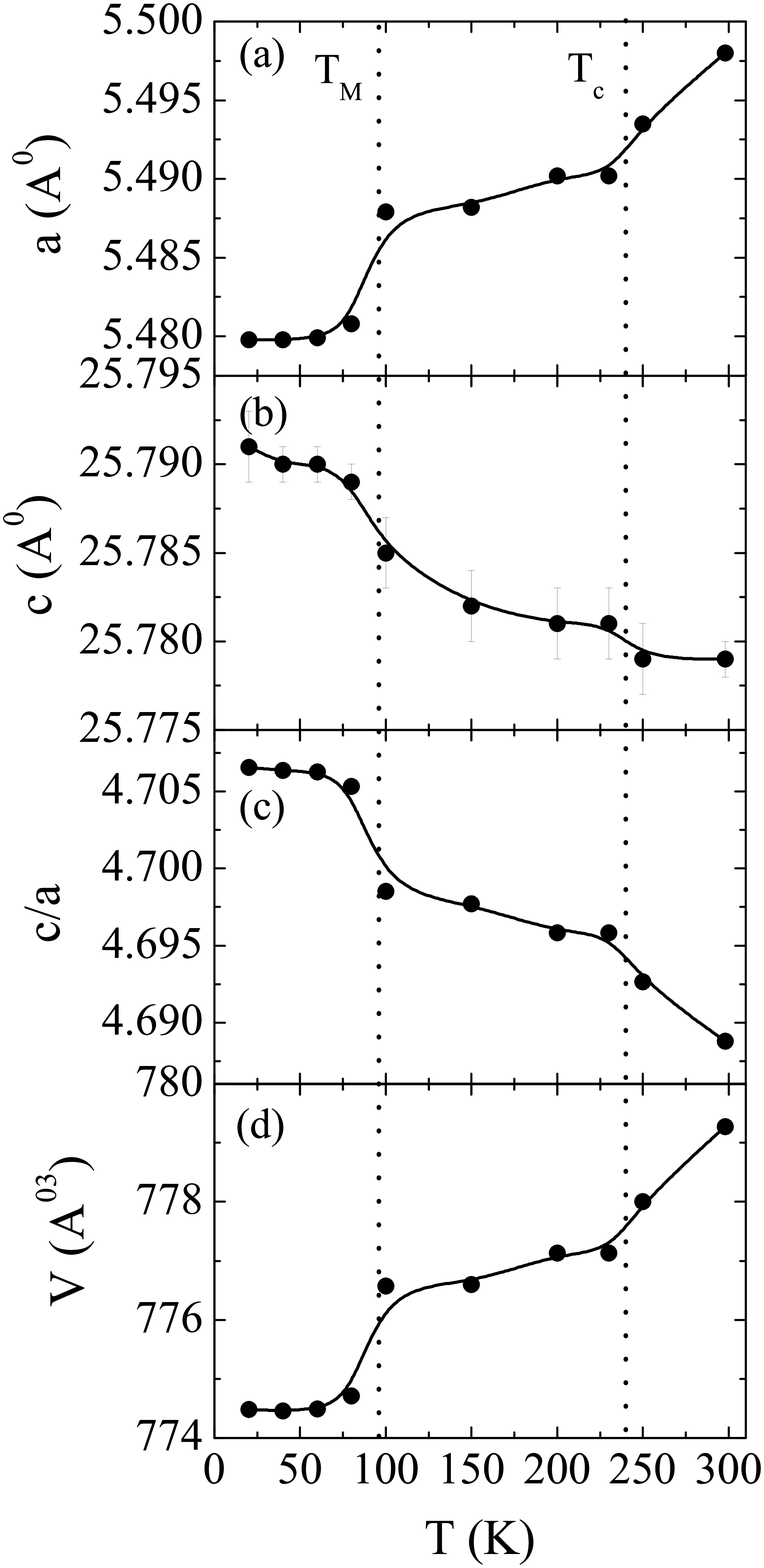}
	\caption{Temperature variation in lattice parameters (a) $a$, (b) $c$, (c) $c/a$ and (d) $V$ are shown for Sr$_2$IrO$_4$. Lines are guide to eyes.}
	\label{fig:Fig6}
\end{figure}

\begin{figure}
	\centering
		\includegraphics[width=7cm]{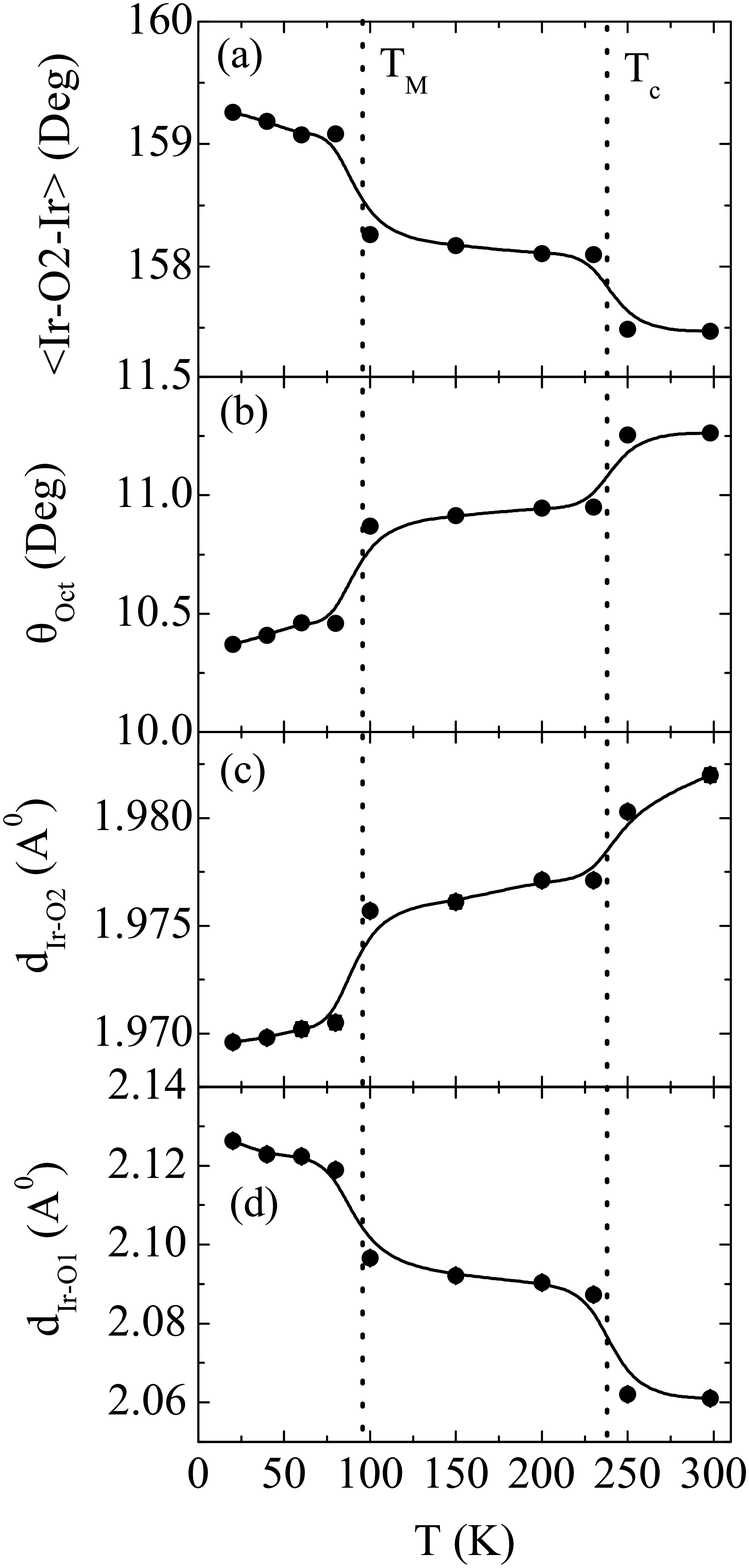}
	\caption{Temperature variation in IrO$_6$ octahedral parameters (a) Ir-O2-Ir bond angel, $<$Ir-O2-O2$>$ (b) octahedral distortion angel, $\theta_{Oct}$, (c) Ir-O2 bond length, $d_{Ir-O2}$ and (d) Ir-O1 bond length, $d_{Ir-O1}$ are shown for Sr$_2$IrO$_4$. Lines are guide to eyes.}
	\label{fig:Fig7}
\end{figure}

To further elucidate this issue we have looked into temperature evolution of bond angel ($<>$) and bond length ($d$) of IrO$_6$ octahedra. Figs. 7a, 7b, 7c and 7d show that with decreasing temperature the basal bond angel $<$Ir-O2-Ir$>$ increases, rotation angel ($\theta_{Oct}$) of IrO$_6$ octahedra around $c$ axis decreases, basal bond length $d_{Ir-O2}$ decreases and apical bond length $d_{Ir-O1}$ increases. It is observed that similar to unit cell parameters in Fig. 6, the bond angel and length also exhibit slope change around $T_c$ and $T_M$. While there is no structural phase transition across the whole temperature range, the changes in slope in structural parameters in Figs. 6 and 7 is intriguing. This structural behavior in Figs. 6 and 7 could be related to temperature evolution of magnetization ($M_{ZFC}$) behavior in Fig. 2. It is commonly accepted fact that magnetism in this material is governed by structural distortion i.e., cooperative rotation of corner shared IrO$_6$ octahedra.\cite{kim2,crawford} The Sr$_2$IrO$_4$, which is isostructural with La$_2$CuO$_4$, is a $J_{eff}$ = 1/2 Mott insulator with single unpaired electron in spin orbit coupled $J_{eff}$ = 1/2 state. The ground state of Sr$_2$IrO$_4$ is expected to be a Heisenberg like magnetic state with AFM spin ordering.\cite{jack} However, a weak ferromagnetic state is established with onset temperature $T_c$ $\sim$ 238 K (Fig. 2), which is believed to be a result of canted antiferromagnetic ordering induced by DM type antisymmetric exchange interaction.\cite{crawford} The DM interaction mainly arises in magnetic materials without inversion symmetry, and favors a spin canting in otherwise AFM system resulting in weak ferromagnetism. The distortion of IrO$_6$ octahedra in Sr$_2$IrO$_4$ breaks the required inversion symmetry of Ir-O2-Ir network. A dominant SOC effect ($\sim$ 0.4 eV) in Sr$_2$IrO$_4$ also contribute in strengthening the DM interaction in this compound.

The theoretical calculation\cite{jack} has, in fact, predicted an intrinsic connection between structural parameters and magnetic ordering. The study shows spin canting angel ($\phi$) and octahedral distortion angel ($\theta_{Oct}$) are somehow connected, and their relation depends on lattice parameters. For $c < a$, the angle $\phi$ is more than the angel $\theta_{Oct}$, however, $\phi$ decreases with increasing $c/a$ ratio, being equal for cubic structure. The decreasing trend of spin canting angel $\phi$, implies a more moment cancellation hence reduced FM component of moment. The temperature evolution of $c/a$ for Sr$_2$IrO$_4$ in Fig. 6c shows a drastic increase in $c/a$ across $T_M$. The similar behavior is also evident in Fig. 7b where the octahedra rotation angel $\theta_{Oct}(T)$ exhibits a decrease with decreasing temperature. These observations are suggestive of weakening of DM interaction, and fairly explains the evolution of magnetization at low temperature where $M_{ZFC}$ shows downfall below $T_M$ (Fig. 2).       

\begin{figure}
	\centering
		\includegraphics[width=8cm]{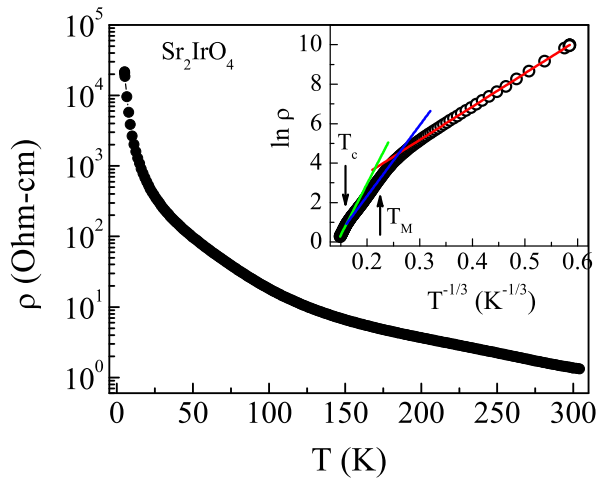}
		\caption{(Color online) Temperature variation in resistivity has been shown for Sr$_2$IrO$_4$ in semilogarithmic plot. Inset shows plotting of same data in form of $\ln \rho$ vs $T^{-1/3}$. Straight lines are fitting of data following Eq. 1. Vertical arrows mark for $T_c$ and $T_M$.}
	\label{fig:Fig8}
\end{figure}

\subsection{Electronic transport study}
To understand the interesting insulating phase in Sr$_2$IrO$_4$, we measured resistivity ($\rho$) as a function of both temperature as well as magnetic field. Fig. 8 shows that at room temperature resistivity is very low with value around 1 $\Omega$-cm. With decreasing temperature the $\rho(T)$ increases exhibiting an insulating behavior,\cite{ge,kini} however, the increase is very drastic at low temperature where resistivity increases by about five orders. We have also measured $\rho(T)$ in presence of applied field of 80 kOe. In presence of magnetic field the qualitative feature of $\rho(T)$ does not change but there is small decrease in $\rho$ where the effect is prominent at lower temperatures (discussed later). This thermally activated conduction behavior has been described using Mott's 2-dimensional (\textit{D}) variable range hopping (VRH) model:\cite{mott}

\begin{eqnarray}
\rho = \rho_0 \exp\left[\left(\frac{T_0}{T}\right)^{1/3}\right]	
\end{eqnarray}

where $T_0$ is the characteristic temperature and can be expressed as:

\begin{eqnarray}
	T_0 = \frac{21.2}{k_BN(E_F)\xi^3}
\end{eqnarray}

\begin{table}
\caption{\label{label} Temperature range, fitting parameter $T_0$ (Eq. 1) and localization length $\xi$ calculated using Eq. 2 are given for Sr$_2$IrO$_4$.}
\begin{ruledtabular}
\begin{tabular}{cccc}
Temperature range (K) &$T_0$ (K) &$\xi$ (\AA)\\
\hline
300 - 240 &1.44 $\times$ 10$^5$ &3.04\\
240 - 70 &4.68 $\times$ 10$^4$ &4.42\\
40 - 5 &4.82 $\times$ 10$^3$ &9.44\\
\end{tabular}
\end{ruledtabular}
\end{table}
  
where $k_B$ is the Boltzmann constant, $N(E_F)$ is the density of states (DOS) at Fermi surface and $\xi$ is the localization length. The inset of Fig. 8 shows straight line fitting of $\rho(T)$ data following Eq. 1. Interestingly, our data can be fitted in three different temperature ranges, and the parameter $T_0$ has been estimated from the respective fittings. Now taking the electronic coefficient of specific heat $\gamma$ = 2 mJ K$^{-2}$ mole$^{-1}$,\cite{kini,carter} we have calculated $N(E_F)$, $\gamma$ = $\pi^2k_B^2V_mN(E_F)$/3, where $V_m$ is the molar volume of Ir. The $N(E_F)$ comes out very low (6 $\times$ 10$^{28}$ eV$^{-1}$ m$^{-3}$) which can be assumed constant across whole temperature as it shows throughout insulating behavior. The so obtained $N(E_F)$ has been used to calculate $\xi$ using Eq. 2. The values are given in Table III. The obtained $T_0$ compares well with other insulating oxide materials.\cite{kini,jender} The $\xi$, which is of the order of planned lattice constant $a$ (Table I), increases in lower temperature range where $\rho$ shows higher values which looks counterintuitive, yet, considering lower thermal energy this increase in resistivity is justified. Moreover, Fig. 8 and Table III show that temperature range for VRH changes around 240 K ($\sim$ $T_c$) and 70 K ($<$ $T_M$), though at low temperature the range is extended up to 40 K below which $\rho(T)$ exhibits steep increase. These behavior is somehow related with both magnetization and structural parameters as they exhibit sudden changes around $T_c$ and $T_M$ (see Figs. 2, 6 and 7). Given that 2$D$ charge transport in this layered tetragonal material the sudden changes of slope in lattice parameter $a$ and basal bond angel $<$Ir-O2-Ir$>$ around $T_c$ and $T_M$ would imply a certain modification in charge hopping which is realized through change in parameter $\xi$ (Table III). It can be noted that the apical bond angle $<$Ir-O1-Ir$>$ does not go through a change retaining value at 180$^o$. Nonetheless, it is important that the three independent measurements i.e., magnetization, structural and resistivity show reasonable agreement which rather authenticates our study.

To understand the electron transport behavior in further detail, we have measured isothermal resistivity as a function of magnetic field up to field of 80 kOe. Fig. 9 shows normalized magnetoresistance (MR) expressed as, $\Delta \rho/\rho(0)$ = $\left[\rho(H) - \rho(0)\right]/\rho(0)$ for Sr$_2$IrO$_4$ at few representative temperatures. As evident in figure, the calculated MR is not very high and at all temperatures its value is negative i.e., resistance decreases in presence of magnetic field in agreement with previous study.\cite{ge} However, the characteristic nature as well as magnitude of MR changes with temperature. For instance, with lowering in temperature the magnitude of MR increases, reaching its value about 4.6\% at 5 K and in magnetic field of 80 kOe. At higher temperatures the MR shows sharp fall in low field regime and then decreases very slowly. In low temperature at 5 K, in contrast, MR does not show any substantial decrease up to field around 16 kOe. Above 16 kOe, it decreases at reasonably constant and faster rate up to heighest measuring field of 80 kOe.

\begin{figure}
	\centering
		\includegraphics[width=8cm]{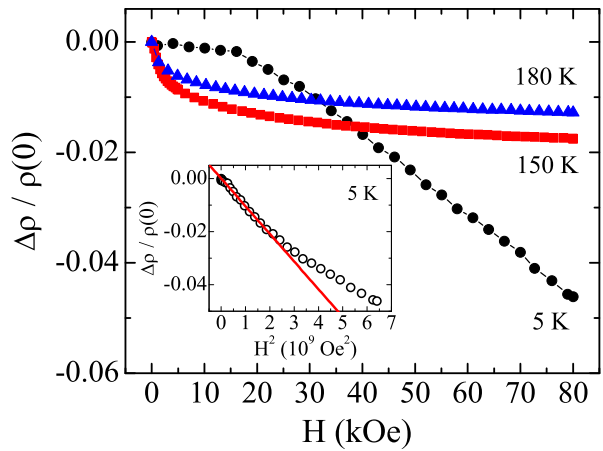}
	\caption{(Color online) Magnetoresistance as a function of magnetic fields are shown for Sr$_2$IrO$_4$ at few representative temperatures. Inset shows quadratic dependence of magnetoresistance on field. Straight line shows fitting of data following Eq. 2.}
	\label{fig:Fig9}
\end{figure}

The MR in VRH regime for disordered materials had drawn interest for long time. In particular, the nature of MR in present material with sizable SOC effect is an interesting subject because a `weak antilocalization' effect induced positive MR has been observed for systems with strong SOC effect such as, Bi$_2$Se$_3$,\cite{chen} Bi$_2$Te$_3$,\cite{he} Au covered Mg film,\cite{berg} even in Ir-based Na$_2$IrO$_3$ films\cite{jender}. In case of Sr$_2$IrO$_4$, a negative MR has been evidenced in bulk sample\cite{ge} while in thin film the MR shows a sign change from positive to negative with temperature around 90 K.\cite{ravi} The negative MR at low temperature in VRH dominated conduction process is commonly understood in picture of `weak localization' effect. This can be interpreted in terms of quantum interference (QI) effect which is largely viewed as quantum correction in the classical Drude equation for conductivity. In case of hopping mediated conduction, the probability of hopping between two suitable sites separated by a distance $R_{hop}$ depends on interference of connecting paths. The effect of magnetic field leads to a destruction of QI effect producing a negative MR.\cite{mott} However, the theoretical calculations employing two different approaches have shown two different dependence of MR on applied field. Nguyen, Spivak, and Shklovskii\cite{nguyen} considered the approach of averaging the logarithm of conductivity over many random paths, and they found the effect of magnetic field is linear. On contrary, a quadratic field dependence of MR was obtained by Sivan, Entin-Wohlman and Imry\cite{imry} who applied a method of critical path analysis to the same problem. Nonetheless, independent of field dependence the negative MR in VRH dominated conduction has been shown as a result of QI effect.

The inset of Fig. 9 shows MR at 5 K follows a quadratic field dependence in lower field regime up to field about 46 kOe. Note, that the MR at higher temperatures (150 and 180 K) does not exhibit such field dependence. This agreement shows QI effect is present in this compound. It can be mentioned that negative MR with quadratic dependence has been observed for many insulating disordered materials such as, $n$-type CdSe,\cite{cdse} indium oxide films,\cite{ino} $n$-type GaAs,\cite{gasa} etc. However, the observation of negative MR in Sr$_2$IrO$_4$ with arguably high SOC is quite surprising while a positive MR is generally evidenced in strong SOC based materials. It is worth noting that a positive MR is observed at low temperature in thin film of Sr$_2$IrO$_4$.\cite{ravi} This implies, even though 2\textit{D} hopping based charge transport is observed in present bulk Sr$_2$IrO$_4$ (Fig. 8), a real 2\textit{D} localization of charges, as in thin films, is required for `weak antilocalization' induced positive MR effect. Nonetheless, more investigations employing films with varying thickness and theoretical calculations are required to understand the MR behavior in Ir-based oxides.
             
\section{Conclusion}
In conclusion, we have prepared polycrystalline sample of layered 5\textit{d} oxide material Sr$_2$IrO$_4$. Analysis of room temperature XRD data shows crystallographically the material adopts a tetragonal structure with \textit{I4$_1$/acd} symmetry where the IrO$_6$ octahedra exhibits a rotation around $c$ axis by angel 11.26$^o$. This is in conformity with previous studies. DC magnetization measurement shows a transition from paramagnetic to weak ferromagnetic state around 238 K, which is also confirmed by Arrott plot. At low temperature below 95 K, a sharp and continuous decrease in magnetization is observed. The analysis of temperature dependent XRD data shows a continuous change in structural parameters, however, a sudden change in parameters is observed around 95 K which is ascribed to observed magnetic behavior. Resistivity measurements show a throughout insulating behavior where the resistivity shows a drastic increase at low temperature increasing by about five orders. The nature of charge transport in all temperatures has been found to follow Mott's 2\textit{D} VRH model, yet three different temperature ranges have been found where the localization length changes accordingly. A negative MR is observed in all temperatures, and the quadratic field dependence of MR at low temperature suggests a presence of quantum interference effect.      

\section{Acknowledgment}   
We are thankful to AIRF, JNU for low temperature XRD measurements and Manoj Pratap Singh for the help in measurements. We acknowledge UGC-DAE CSR, Indore for magnetization and resistivity measurements, and Kranti Kumar and Sachin Kumar for the helps in measurements.


\begin{thebibliography}{}
\bibitem{kim1} B. J. Kim, Hosub Jin, S. J. Moon, J.-Y. Kim, B.-G. Park, C. S. Leem, Jaejun Yu, T.W. Noh, C. Kim, S.-J. Oh,
J.-H. Park, V. Durairaj, G. Cao, and E. Rotenberg, Phys. Rev. Lett. \textbf{101}, 076402 (2008).
\bibitem{kim2} B. J. Kim, H. Ohsumi, T. Komesu, S. Sakai, T. Morita, H. Takagi, T. Arima, Science \textbf{323}, 1329 (2009).
\bibitem{balent} W. Witczak-Krempa, G. Chen, Y. B. Kim, and L. Balents, Annu. Rev. Condens. Matter Phys. \textbf{5}, 57 (2014).
\bibitem{naka} S. Nakatsuji, Y. Machida, Y. Maeno, T. Tayama, T. Sakakibara, J. van Duijn, L. Balicas, J. N. Millican, R. T. Macaluso, and Julia Y. Chan, Phys. Rev. Lett. \textbf{96}, 087204 (2006).
\bibitem{shapiro} M. C. Shapiro, S. C. Riggs, M. B. Stone, C. R. de la Cruz, S. Chi, A. A. Podlesnyak, and I. R. Fisher, Phys. Rev. B \textbf{85}, 214434 (2012).
\bibitem{yogesh} Yogesh Singh, S. Manni, J. Reuther, T. Berlijn, R. Thomale, W. Ku, S. Trebst, and P. Gegenwart, Phys. Rev. Lett. \textbf{108}, 127203 (2012).
\bibitem{okamoto} Y. Okamoto, M. Nohara, H. Aruga-Katori, and H. Takagi, Phys. Rev. Lett. \textbf{99}, 137207 (2007).
\bibitem{moon} S. J. Moon, H. Jin, K.W. Kim, W. S. Choi, Y. S. Lee, J. Yu, G. Cao, A. Sumi, H. Funakubo, C. Bernhard, and T.W. Noh, Phys. Rev. Lett. \textbf{101}, 226402 (2008).
\bibitem{jack} G. Jackeli1, and G. Khaliullin, Phys. Rev. Lett. \textbf{102}, 017205 (2009).
\bibitem{senthil} Fa Wang and T. Senthil, Phys. Rev. Lett. \textbf{106}, 136402 (2011).
\bibitem{yang} Yang Yang, W-S. Wang, J-G. Liu, H. Chen, J.-H. Dai, and Q.-H. Wang, Phys. Rev. B \textbf{89}, 094518 (2014).
\bibitem{bahr} S. Bahr, A. Alfonsov, G. Jackeli, G. Khaliullin, A. Matsumoto, T. Takayama, H. Takagi, B. B\"{u}chner, and V. Kataev, Phys. Rev. B \textbf{89} 180401 (2014).
\bibitem{crawford} M. K. Crawford, M. A. Subramanian, and R. L. Harlow, J. A. Fernandez-Baca, Z. R. Wang and D. C. Johnston, Phys. Rev. B \textbf{49}, 9198 (1994).
\bibitem{ye} Feng Ye, Songxue Chi, B. C. Chakoumakos, J. A. Fernandez-Baca, T. Qi, and G. Cao, Phys. Rev. B \textbf{87}, 140406 (2013).
\bibitem{ge} M. Ge, T. F. Qi, O. B. Korneta, D. E. De Long, P. Schlottmann, W. P. Crummett, and G. Cao, Phys. Rev. B \textbf{84}, 100402, (2011).
\bibitem{korneta} O. B. Korneta, Tongfei Qi, S. Chikara, S. Parkin, L. E. De Long, P. Schlottmann, and G. Cao, Phys. Rev. B \textbf{82}, 115117 (2010).
\bibitem{chikara} S. Chikara, O. Korneta, W. P. Crummett, L. E. DeLong, P. Schlottmann, and G. Cao, Phys. Rev. B \textbf{80}, 140407 (2009).
\bibitem{ge1} M. Ge, S. Tan, Y. Huang, L. Zhang, W. Tong, L. Pi, and  Y. Zhang, J. Magn. Magn. Mater. \textbf{345}, 13 (2013).
\bibitem{young} R. A. Young, A. Sakthivel, T. S. Moss and C. O. Paiva-Santos, \textsl{Users guide to program DBWS-9411, Atlanta: Georgia Institute of Technology} (1994).
\bibitem{cao} G. Cao, J. Bolivar, S. McCall, and J. E. Crow, andR. P. Guertin, Phys. Rev. B \textbf{57}, 11039 (1998).
\bibitem{kini} N. S. Kini, A. M. Strydom, H. S. Jeevan, C. Geibel and S. Ramakrishnan, J. Phys.: Condens. Matter \textbf{18}, 8205 (2006).
\bibitem{calder} S. Calder, G.-X. Cao, M. D. Lumsden, J. W. Kim, Z. Gai, B. C. Sales, D. Mandrus, and A. D. Christianson, Phys. Rev. B \textbf{86}, 220403, (2012).
\bibitem{arrott} A. Arrott, Phys. Rev. \textbf{108}, 1394 (1957).
\bibitem{rhodes} P. Rhodes and E. P. Wohlfarth, Proc. R. Soc. London 273, 247 (1963).
\bibitem{mott} N. Mott, \textit{Conduction in Non-Crystalline Materials}, Clarendon Press, Oxford (1993).
\bibitem{carter} S. A. Carter, B. Batlogg, R. J. Cava, J. J. Krajewski, W. F. Peck, Jr., and L. W. Rupp, Jr., Phys. Rev. B \textbf{51}, 17184 (1995).
\bibitem{jender} M. Jenderka, J. Barzola-Quiquia, Z. Zhang, H. Frenzel, M. Grundmann, and M. Lorenz, Phys. Rev. B \textbf{88}, 045111 (2013).
\bibitem{chen} J. Chen, H. J. Qin, F. Yang, J. Liu, T. Guan, F. M. Qu, G. H. Zhang, J. R. Shi, X. C. Xie, C. L. Yang, K. H. Wu, Y. Q. Li, and L. Lu, Phys. Rev. Lett. \textbf{105}, 176602 (2010).
\bibitem{he} H-T. He, G. Wang, T. Zhang, I-K. Sou, G. K. L. Wong, and J-N. Wang, Phys. Rev. Lett. \textbf{106}, 166805 (2011).
\bibitem{berg} G. Bergman, Phys. Rev. Lett. \textbf{48}, 1046 (1982).
\bibitem{ravi} J. Ravichandran, C. R. Serrao, D. K. Efetov, D. Yi, R. Ramesh, and P. Kim, arXiv:1312.7015.
\bibitem{nguyen} V. L. Nguyen, B. Z. Spivak, and B. I. Shklovskii, Zh. Eksp. Teor. Fiz. \textbf{89}, 1770 (1985). [Sov. Phys. — JETP \textbf{62}, 1021 (1985)].
\bibitem{imry} U. Sivan and O. Entin-Wohlman, and Y. Imry, Phys. Rev. Lett. \textbf{60}, 1566 (1988).
\bibitem{cdse} Y. Zhang and M. P. Sarachik, Phys. Rev. B \textbf{43}, 7212 (1991).
\bibitem{ino} O. Faran and Z. Ovadyahu, Phys. Rev. B \textbf{38}, 5457 (1988).
\bibitem{gasa} F. Tremblay, M. Pepper, D. Ritchie, D. C. Peacock, J. E. F. Frost, and G. A. C. Jones, Phys. Rev. B \textbf{39}, 8059 (1989).
\end{thebibliography}
\end{document}